\documentclass[12pt,a4paper]{article}
\usepackage[T1]{fontenc}
\usepackage[utf8]{inputenc}
\usepackage{authblk}
\usepackage{amsmath,amssymb,epsfig,cite,dsfont}
\usepackage{array,multirow}
\usepackage {color}
\numberwithin{equation}{section}

\usepackage[colorlinks=true
,urlcolor=blue
,anchorcolor=blue
,citecolor=blue
,filecolor=blue
,linkcolor=blue
,menucolor=blue
,pagecolor=blue
,linktocpage=true
,pdfproducer=medialab
,pdfa=true
]{hyperref}

\advance\hoffset by -1.1truecm
\advance\voffset by -1.0truecm
\newcommand{\be}{\begin{equation}}
\newcommand{\ee}{\end{equation}}
\newcommand{\bea}{\begin{eqnarray}}
\newcommand{\eea}{\end{eqnarray}}

\usepackage{mathrsfs}
\usepackage{authblk}

\numberwithin{equation}{section}
\usepackage{float}
\restylefloat{figure}

\usepackage{epsfig}
\usepackage{slashed}
\usepackage[titletoc,title]{appendix}

\def\be{\begin{equation}}
\def\ee{\end{equation}}
\def\bea{\begin{eqnarray}}
\def\eea{\end{eqnarray}}
\begin{document}

\title{\vspace{0cm}\begin{flushleft} \bf { Supersymmetry: Boundary Conditions and Edge States
\linethickness{.05cm}\line(1,0){433}
}\end{flushleft}}
\author[a]{\bf Nirmalendu~Acharyya\footnote{nirmalendu@cts.iisc.ernet.in}}
\author[b]{\bf  Manuel Asorey\footnote{asorey@unizar.es}}
\author[c,d]{\bf A.~P.~Balachandran\footnote{balachandran38@gmail.com}}
\author[a]{\bf Sachindeo~Vaidya\footnote{vaidya@cts.iisc.ernet.in}}
\affil[a]{ \small Centre for High Energy Physics,  Indian Institute of Science, Bangalore-560012,
India}
\affil[b]{ \small Departamento de F\'isica Te\'orica,
Universidad de Zaragoza,
E-50009 Zaragoza, Spain}
\affil[c]{ Department of Physics, Syracuse University, Syracuse, N. Y. 13244-1130, USA}
\affil[d]{ Departamento de Fisica, Universidad de los Andes, Bogot\'a, Colombia}
\date{}
\maketitle
\abstract{When spatial boundaries are inserted, SUSY can be broken. 
We show that in an $\mathcal{N}=2$ supersymmetric theory,  all the boundary conditions allowed by self-adjointness of the Hamiltonian break $\mathcal{N}=2$ SUSY while only a few of these boundary conditions preserve  $\mathcal{N}=1$ SUSY. We  also show that for a subset of the boundary conditions compatible with $\mathcal{N}=1$ SUSY, there exist fermionic ground states which are localized near the boundary. 
}

\section{Introduction}

Although supersymmetry (SUSY) as a fundamental theory has eluded experimental evidence to date, there has been a recent revival of interest in the subject because 
it emerges naturally as an effective theory describing the quantum phase transition at the boundary of topological superconductors \cite{grover}.

All real physical systems available for experiments are of finite size and with spatial boundaries, which in general reduce the symmetries of the system. Hence it is eminently reasonable to ask if the SUSY of a $(d+1)$-dimensional finite size system (like the  topological superconductor) can be obtained by the consistent truncation of a parent SUSY system in full $(d+1)$-dimensional Minkowski spacetime.
Although one might expect that insertion of spatial boundaries generically breaks SUSY, we will show that there are certain boundary 
conditions which do preserve supersymmetry partially. Discussions of boundary conditions in this context assume significance, and a clear classification of such boundary conditions is required. The presence of boundaries, on the other hand, naturally leads to the the question of edge states, which, if extant, play a vital role in  the physics at the boundary \cite{Qi}.

 Boundary conditions in supersymmetric theories have been studied in details (for example see ~\cite{Belyaev:2008xk} and references there in). We consider
this problem from a different perspective. We show that self-adjoint domains of the Hamiltonian are enough
to obtain the boundary conditions which preserve (or break) supersymmetry.  Further, our main objective is to
show the existence of edge states in a supersymmetric theory, which will be relevant in the physics of
the newly discovered supersymmetric phase in topological superconductors. For this purpose, a simplified
treatment of a non-interacting scalar-fermion model is sufficient. As discussed in~\cite{Asorey:2006pr, Asorey:2008xt, Asorey:2011zza, Asorey:2015lja}, for the above
mentioned model, it is not difficult to see to that a supersymmetric variation in the bulk gives boundary terms which vanish only when
Dirichlet or Neumann boundary conditions are chosen for the scalar. It is well-known that with
Dirichlet or Neumann boundary conditions there are no scalar edge localized states~\cite{aim}. Nevertheless,
we show that there can still be fermionic edge states which do not break supersymmetry.

In \cite{grover} it was shown that in the phase that breaks SUSY spontaneously, there are edge states on the surface (i.e. the boundary) of 
the superconductor. However, it is not obvious whether such edge states exist without breaking SUSY.  We will 
investigate the existence of such edge states when the boundary conditions can be chosen to preserve (some) supersymmetry.  As we will show, 
such ``SUSY preserving" edge states do exist, and the ground 
states in such theories are particularly interesting.

Our focus in this article will be on the insertion of a spatial boundary $\partial M$ in $(d+1)$-dimensional Minkowski space, in such a manner that the resulting space continues to be a $(d+1)$-dimensional manifold $M$ with a boundary $\partial M$ (which can be curved in general). The boundary conditions on the (scalar and spinor) fields on $M$ cannot 
be chosen arbitrarily. They are obtained by demanding that the scalar and Dirac Hamiltonians ($H_s$ and $H_D$ respectively) be self-adjoint. Of these boundary conditions, we expect that only a subset will preserve supersymmetry (at least  partially), while generic boundary conditions will break supersymmetry completely.

For $H_s$ to be self-adjoint, it is necessary that the scalar Laplacian $(-\nabla^2+m^2)$ be self-adjoint \cite{aim}. Then, if we demand locality of boundary conditions, the domain $ \mathcal{D}_{H_s}=\mathcal{D}_{H_s^\ast}$ of $H_s$ contains all $\Phi \in L^2(M)$ satisfying 
\begin{eqnarray}
\label{allowed_scalar_bc1}\left.
\begin{array}{rcl}
\Big[\Phi(x)+ i\partial_n \Phi(x)\Big] &=& U_B(x) \Big[\Phi(x)-i \partial_n \Phi(x)\Big], \\
 U_B^{\dagger}(x) U_B(x)&=&\mathbb{I}, \,\,\,\, \quad x \in \partial M
\end{array}\right. 
\end{eqnarray}
where $\hat{n}$ is the outward normal, $\partial_n \equiv \hat{n}\cdot\vec{ \nabla}$ is the normal derivative at $\partial M$ and $U_B(x)$ is a unitary operator on $\Phi(x)$ (if $\Phi(x)$ is $N$ component, $U_B(x) \in U(N)$).

For  the choices $U_B(x)=-\mathbb{I}_{N\times N}$ and $U_B(x)=\mathbb{I}_{N\times N}$, we get the Dirichlet and Neumann boundary conditions respectively. Other choices for $U_B(x)$ give more general boundary conditions:
\begin{equation}
\Big[\partial_n \Phi  - \kappa \Phi \Big](x)=0, \,\,\,\, \kappa=i \left({\mathbb{I}+U_B}\right)^{-1}\left({\mathbb{I}-U_B}\right), \,\,\,\, \kappa(x)^\dagger= \kappa(x)
\label{robinbc}
\end{equation}
whenever $U_B$ does not have unit eigenvalues.

To discuss the self-adjointness of $H_F\equiv i\gamma^0\gamma^j \partial_j-m\gamma^0$, we start by defining two chiralities (on the boundary) for the Dirac spinors $\Psi$:
\begin{equation}
\Psi_\pm \equiv  \frac{1}{2} \left(1\pm \gamma^0 \vec{\gamma}\cdot\hat{n}\right) \Psi. 
\end{equation}
The $\gamma$-matrices here obey
\begin{equation}
\{\gamma^\mu, \gamma^\nu\}=2 \eta^{\mu\nu}, \quad  \gamma^{0\dagger} =\gamma^0, \quad  \gamma^{j\dagger}=-\gamma^j, 
\end{equation}
where $\mu,\nu=0,1\cdots d$, $j=1,2\cdots d$, and $\eta^{\mu\nu}=\textrm{diag}(1,-1,-1,\cdots-1)$.

The essential self-adjointness of $H_F$ requires that the domains
$\mathcal{D}_{H_F}$ and  $\mathcal{D}_{H_F^\ast}$ coincide. If $M$ is compact, the most general self-adjoint extension fulfilling local boundary condition is given by
$\Psi \in W^{1,2}(M) \otimes {\mathbb C}^N$ satisfying  \cite{asorey, aim}
\begin{eqnarray}\label{allowed_spinor_bc1}
 \left(\Psi_+ - U_F \gamma^0 \Psi_- \right)_{\partial M}=0,\,\,\, U^{\dagger}_F U_F=1,\,\,\, [U_F, \gamma^0 \vec{\gamma}\cdot \hat{n}]=0.
\end{eqnarray}

We now analyse these general observations in various dimensions.

\section{(1+1)-dimensions}
In the full (1+1)-dimensional Minkowski spacetime, the simplest theory of a complex scalar $\Phi$ (with the number of components $N=1$) and a Dirac fermion  $\Psi$ is  $\mathcal{N}=2$ supersymmetric \cite{hori}. It is described by the action
\begin{equation}\label{action456}
 S= \int_{-\infty}^{\infty} dx_0 \int_{-\infty}^\infty dx_1 \left(\mathcal{L}_{S} +\mathcal{L}_{D}\right)  
\end{equation}
where
\begin{eqnarray}
&&\mathcal{L}_{S} =\frac{1}{2}\left(\partial^\mu\Phi^\ast \partial_\mu \Phi -m^2 \Phi^\ast \Phi\right), \\
&& \mathcal{L}_{F}=\frac{1}{2}\left(i \bar{\Psi} \gamma^{\mu} \partial_\mu \Psi -m\bar{\Psi}\Psi\right).
\end{eqnarray}

 The SUSY transformations are
\begin{eqnarray}\label{susy_transformations}
\begin{array}{lll}
\delta \Phi =\bar{ \epsilon} \Psi, \quad \quad \delta \Psi = -i \gamma^\mu \epsilon \partial_\mu\Phi -m \epsilon \Phi, \\
 \delta \Phi^\ast =\bar{\Psi} \epsilon, \quad\quad \delta \bar{\Psi} = i  \bar{\epsilon}\gamma^\mu \partial_\mu\Phi^\ast -m \bar{\epsilon} \Phi^\ast, 
\end{array}
\end{eqnarray}
with
\begin{eqnarray}
\epsilon = \left(\begin{array}{ll}
\epsilon_1\\
\epsilon_2
\end{array}\right)\nonumber
\end{eqnarray}
where $ \epsilon_i$'s are Grassmann constants and $\bar{\epsilon} = \epsilon^\dagger \gamma^0$.

We consider the same system in a (1+1)-dimensional manifold $M$ with spatial boundary $\partial M$:
\begin{equation}
M=	\{(x^0, x^1): \quad x^1 \leq 0\}.
\end{equation}
 The action is given by 
\begin{equation}\label{action}
S= \int_{-\infty}^{\infty} dx_0 \int_{-\infty}^0 dx_1 \left(\mathcal{L}_{S} +\mathcal{L}_{D}\right) +S_{B}  
\end{equation}
 and $S_B$ are the boundary terms on $\partial M$ as in \cite{Asorey:2006pr, Asorey:2008xt}: 
\begin{eqnarray}
&& S_B=\frac{1}{4}\int_{-\infty}^\infty dx_0\left(\Phi^\ast \partial_n\Phi+(\partial_n\Phi^\ast
)\Phi-i\bar{\Psi}\vec{\gamma}\cdot\hat{n}\Psi\right)_{\partial M}.
\end{eqnarray}
The boundary terms are analogous to the Gibbons-Hawking term. Its goal is to give rise to local equations of motion independently of boundary conditions on the fields.

The boundary conditions (\ref{allowed_scalar_bc1}) and (\ref{allowed_spinor_bc1}) are imposed on $\Phi$ and $\Psi$ at the boundary points $x^1=0$.  Out of this family of allowed boundary conditions, which ones are consistent with the SUSY transformations (\ref{susy_transformations})?

The SUSY transformation $\delta \Phi$ and $\delta \Psi$ must obey  (\ref{allowed_scalar_bc1}) and and (\ref{allowed_spinor_bc1}) on the boundary. The variation of the scalar field on the boundary leads  to
\begin{eqnarray}\label{trans_1d}
\Big[(1-U_B)\bar{ \epsilon} \Psi +i (1+U_B)\bar{ \epsilon} (\partial_1\Psi )\Big]_{\partial M}=0,
\end{eqnarray}
 $U_B$ in this case being a phase.  The variation of $\Psi$ on the boundary yields
\begin{eqnarray}
\Big[- i (\gamma^\mu \epsilon)_+ \partial_\mu \Phi - m \epsilon_+ \Phi \Big]_{\partial M}= \Big[-i U_F \gamma^0 (\gamma^\mu \epsilon)_- \partial_\mu \Phi - m U_F \gamma^0 \epsilon_- \Phi\Big]_{\partial M}
\end{eqnarray}
which leads to
\begin{equation}\label{trans_1df}
\Big[i \gamma^\mu \epsilon_- \partial_\mu \Phi+ m \epsilon_+ \Phi \Big]_{\partial M}=\Big[ i  \gamma^0 \gamma^\mu  U_F \epsilon_+ \partial_\mu \Phi+ m U_F \gamma^0 \epsilon_- \Phi\Big]_{\partial M}.
\end{equation}
It can be easily checked that (\ref{trans_1d}) and(\ref{trans_1df}) are incompatible if $U_B \neq \pm1$ (Dirichlet or Neumann).
 \textit{Therefore if Robin boundary conditions are imposed on scalars ($\kappa\neq 0$ or $\kappa\neq \infty$  where $\kappa$ is defined in (\ref{robinbc})) in the (1+1)-dimensional theory,  $\mathcal{N}=2$ SUSY is completely broken.}

\textbf{Dirichlet and Neumann boundary conditions:}
However Dirichlet and  Neumann boundary conditions on the scalar are consistent with SUSY.  To show that let us consider the  massless and massive cases separately.
\newline 
{\textit{Massless case:}}
If we impose the Dirichlet $(U_B=-1)$ or Neumann $(U_B=1)$ condition on the massless scalar $\Phi$,
the supersymmetry condition (\ref{trans_1df}) leads to 
\begin{eqnarray}
&&\textrm{Dirichlet}: \quad  i \gamma^1 \epsilon_- \partial_1 \Phi\Big|_{\partial M}  =-i \gamma^0\gamma^1 U_F\epsilon_+ \partial_1 \Phi\Big|_{\partial M} \\
&&\textrm{Neumann}:\quad   i \gamma^0 \epsilon_- \partial_0 \Phi\Big|_{\partial M}  =-i \gamma^0\gamma^0 U_F\epsilon_+ \partial_0 \Phi\Big|_{\partial M} 
\end{eqnarray}
which yields the following condition on the SUSY parameter $\epsilon$:
\begin{eqnarray}
&&\textrm{Dirichlet}: \quad  \epsilon_-  =- \gamma^0 U_{F }  \epsilon_+ , \label{relation_UF_1d} \\
&&\textrm{Neumann}:\quad  \epsilon_-  =\gamma^0 U_{F}  \epsilon_+. \label{relation_UF_1n}
\end{eqnarray}
On the other hand, these choices  $U_B = \pm 1$ in (\ref{trans_1d}) give
\begin{eqnarray}
&&\textrm{Dirichlet}: \quad \bar{\epsilon}\Psi\Big|_{\partial M} =0,  \label{dc1}\\
&&\textrm{Neumann}:\quad \bar{\epsilon}\partial_1\Psi\Big|_{\partial M} =0. \label{nc1}
\end{eqnarray}

For the Dirichlet boundary condition on the scalar,  the condition (\ref{dc1}) is trivially satisfied when the boundary condition  (\ref{allowed_scalar_bc1}) and the condition (\ref{relation_UF_1d}) on $\epsilon$  are used.  

For Neumann boundary condition on the scalar, the condition (\ref{nc1}) along with (\ref{relation_UF_1n})) yield a new boundary condition 
\begin{equation}\label{cond_952}
(\partial_1 \Psi)_+\Big|_{\partial M} = -U_F \gamma^0 (\partial_1 \Psi)_-\Big|_{\partial M}.
\end{equation}
However, appearance of this extra boundary condition is not surprising in a supersymmetric theory.  The supercharge $Q$ obeys
\begin{equation}
\{Q,\bar{Q}\} \Psi \propto H_F \Psi. 
\end{equation}
Hence, it is necessary to ensure that $(H_F \Psi)$ is also in the domain of $H_F$. Otherwise SUSY will change the domain of $H_F$. Hence,  we must also impose 
\begin{eqnarray}\label{cond_953}
(H_F \Psi)_+\Big|_{\partial M} = U_F \gamma^0 (H_F \Psi)_-\Big|_{\partial M} 
\end{eqnarray}
which in massless (1+1)-dimensional case reduces to (\ref{cond_952}).

\textit{Therefore, the Dirichlet (or Neumann) boundary condition on massless $\Phi$ is consistent with the supersymmetry transformations and the system is supersymmetric. But owing to the relation (\ref{relation_UF_1d})  (or (\ref{relation_UF_1n})), the system has only $\mathcal{N}=1$ supersymmetry.}
\newline
{\textit{Massive case:}
If the Dirichlet boundary condition ($U_B=-1$) is imposed on the scalar, 
the supersymmetry condition (\ref{trans_1d}) and the boundary condition  (\ref{allowed_spinor_bc1}) lead to 
\begin{equation}
   \epsilon_-  =- \gamma^0 U_{F }  \epsilon_+ . \label{relation_UF_1dm}
\end{equation}
With Dirichlet boundary condition on $\Phi$ and (\ref{relation_UF_1dm}),  it is easy to see from (\ref{trans_1df}) that
\begin{equation}\label{proof_234m}
\delta \Psi _+ = U_F \gamma^0 \delta \Psi _+
\end{equation}
is satisfied. Therefore, this choice of boundary conditions is consistent with SUSY.

If Neumann boundary condition ($U_B=1$) is imposed on the scalar, the supersymmetry condition (\ref{trans_1d}) gives
\begin{equation}\label{eqn_20}
\bar{\epsilon}( \partial_1 \Psi )\Big|_{\partial_M} =0,
\end{equation}
while (\ref{trans_1df}) leads to 
\begin{equation}\label{trans_1dfmass}
\Big[i \gamma^0 \epsilon_- \partial_0 \Phi+ m \epsilon_+ \Phi \Big]_{\partial M}=\Big[ i  \gamma^0 \gamma^0  U_F \epsilon_- \partial_0 \Phi+ m U_F \gamma^0 \epsilon_- \Phi\Big]_{\partial M}.
\end{equation}
In contrast to the massless case, here,  because of the extra mass term in (\ref{trans_1dfmass}), this cannot be made compatible  with (\ref{eqn_20}) just by a condition on $\epsilon$.  However, the two can be made compatible by imposing the further condition  $U_F=U_F^\dagger$. Hence the Neumann boundary condition on the massive scalar in (1+1)-dimension is consistent with SUSY when 
\begin{equation}\label{eqn_24}
  \epsilon_-  = \gamma^0 U_{F }  \epsilon_+, \quad\quad U_F= U_F^\dagger. 
\end{equation}

  \textit{As a result, in the massive (1+1)-dimensional  theory, imposing Dirichlet or Neumann boundary condition on the scalar breaks $\mathcal{N}=2$ SUSY to $\mathcal{N}=1$ SUSY.}

With the choice  $\gamma^0= \sigma_2$ and $\gamma^1=i\sigma_1$, in (1+1)-dimensional massless case and in the massive case with Dirichlet boundary condition, the most general $U_F$ satisfying (\ref{allowed_spinor_bc1}) is 
\begin{equation}\label{U_F_gen_1d}
U_F = \left(\begin{array}{cccc}
e^{i\theta}&0\\
0& e^{i\tilde{\theta}}
\end{array}\right), \quad\quad \theta, \tilde{\theta} \in \mathbb{R}.
\end{equation}
For the massive case with Neumann boundary condition, because of the condition (\ref{eqn_24}), only $\theta= \tilde{\theta}= 0 \textrm{ or } \pi$ are allowed by supersymmetry and hence the only $U_F$'s that preserve SUSY partially are
\begin{equation}
U_F = \pm \left(\begin{array}{cccc}
1&0\\
0& 1
\end{array}\right).
\end{equation}
Using the above in (\ref{relation_UF_1d}), (\ref{relation_UF_1n}), (\ref{relation_UF_1dm}) and (\ref{eqn_24}), we get  
\begin{eqnarray}
&&\textrm{Dirichlet}: \quad \quad\quad\quad \quad \,\,\, \,\,\, \, \epsilon_1 = - i e^{-i \theta} \epsilon_2, \\
&&\textrm{Neumann (massless)}: \quad \epsilon_1 =  i e^{-i \theta} \epsilon_2 \\
&&\textrm{Neumann (massive)}: \quad \,\, \epsilon_1 =  \pm i  \epsilon_2.
\end{eqnarray}
The closure of the SUSY algebra is given by 
\begin{eqnarray}\label{closure}
\left[\delta_{\epsilon} , \delta_{\eta}\right]  = -i ({\epsilon}^\dagger \eta -{\eta}^\dagger \epsilon) \partial_0  + 2m  (\bar{\epsilon} \eta -\bar{\eta} \epsilon) .
\end{eqnarray}

The unbroken $\mathcal{N}=2$ SUSY algebra in (1+1)-dimensions is generated by two supercharges $Q_\pm$: 
\begin{eqnarray}
\{Q_\pm, \bar{Q}_\pm\} = \mathcal{P}_0 \pm \mathcal{P}_1, \quad \{Q_-, \bar{Q}_+\} = \mathcal{Z}, \quad \{Q_+, \bar{Q}_-\} = \bar{\mathcal{Z}}, \quad [P_\mu, \mathcal{Z}]=0,
\end{eqnarray}
where $\mathcal{Z}$ is the central charge.
In the $\mathcal{N}=1$ theory, as the SUSY parameter satisfies (\ref{relation_UF_1d}) or (\ref{relation_UF_1n}) in the massless case and  (\ref{relation_UF_1dm}) or (\ref{eqn_24}) in the massive case, the super charges are
\begin{eqnarray}
&&\textrm{Dirichlet}: \quad\quad\quad\quad\quad\quad\, Q = Q_+ + i e^{i \theta} Q_-, \\
&&\textrm{Neumann (massless)}: \quad  Q = Q_+ - i e^{i \theta} Q_-,\\
&&\textrm{Neumann (massive)}: \quad  \,\,Q = Q_+ \pm i  Q_-,
\end{eqnarray}
satisfying
\begin{eqnarray}\hspace*{-1cm}\left.\begin{array}{lll}
&&\textrm{Dirichlet}: \quad\quad\quad\quad\quad\quad \,\{Q, \bar{Q}\} =2 \mathcal{P}_0- i(e^{-i \theta} \bar{\mathcal{Z}} -e^{i \theta} \mathcal{Z}),  \\
&&\textrm{Neumann (massless)}: \quad \{Q, \bar{Q}\} =2 \mathcal{P}_0+i(e^{-i \theta} \bar{\mathcal{Z}} -e^{i \theta} \mathcal{Z}), \\
&&\textrm{Neumann (massive)}: \quad \,\,\{Q, \bar{Q}\} =2 \mathcal{P}_0\pm i( \bar{\mathcal{Z}} - \mathcal{Z}), \end{array}\right\}\quad [Q, \mathcal{P}_0]=0.
\end{eqnarray}
In (\ref{closure}), the mass term is the central charge contribution (i.e. the massless theory has $\mathcal{Z}=0$). 
In the massless case, this term vanishes and we get the usual $\mathcal{N}=1$ SUSY algebra. But in the massive case, the central charge term can be absorbed by rescaling $\mathcal{P}_0$ and the usual $\mathcal{N}=1$ SUSY algebra can be recovered: 
\begin{eqnarray}\label{modified_algebra}&& \left.\begin{array}{ll}
\textrm{Dirichlet (massive)}: \quad \,\,\tilde{\mathcal{P}}_0 =  \mathcal{P}_0- \frac{i}{2}(e^{-i \theta} \bar{\mathcal{Z}} -e^{i \theta} \mathcal{Z}),\\
\textrm{Neumann (massive)}: \quad\tilde{\mathcal{P}}_0 =  \mathcal{P}_0\pm \frac{i}{2}( \bar{\mathcal{Z}} -\mathcal{Z}),
\end{array}\right. \\
&&\quad \{Q, \bar{Q}\} =2 \tilde{\mathcal{P}}_0, \quad\quad \quad  [Q, \tilde{\mathcal{P}}_0]=0.
\end{eqnarray}

\textit{Hence, if the theory is massless, the $\mathcal{N}=2$  SUSY is broken to  a family (characterized by $\theta$ and $\tilde{\theta}$) of $\mathcal{N}=1$ supersymmetric theories by introducing the boundary with Dirichlet or Neumann boundary condition on the scalar. } 

\textit{On the other hand, if the theory is massive, in presence of a boundary with only the Dirichlet boundary condition on the scalar, breaks $\mathcal{N}=2$ SUSY to a family of $\mathcal{N}=1$ supersymmetric theories.  In case of Neumann boundary conditions, $\mathcal{N}=2$ SUSY is broken to one of the two possible $\mathcal{N}=1$ SUSY theories, depending on the fermionic boundary conditions (i.e. only when $U_F=\pm 1$. For any other choice of $U_F$, SUSY is completely broken). Any other boundary condition on the scalar breaks SUSY completely.} 
}

\subsection{Variation of the action}
One can verify that the above results can be simply re-derived by requiring invariance of the full action. Indeed, the  variation of the action (\ref{action}) under SUSY yields
\begin{eqnarray}
\delta S&=& \frac{1}{4}\int_{\partial M} dx^0 \Big[\Phi^\ast (\bar{\epsilon}\partial_1\Psi)+(\partial_1\bar{\Psi} \epsilon)\Phi-\bar{\Psi}\gamma^1\gamma^0\epsilon \partial_0 \Phi + \bar{\epsilon} \gamma^1 \gamma^0\Psi \partial_0\Phi^\ast\Big]_{x_1=0} \nonumber \\
&& +\frac{i m}{4} \int_{\partial M} dx^0 \Big[ \bar{\Psi}\gamma^1 \epsilon \Phi - \bar{\epsilon}\gamma^1 \Psi \Phi^\ast\Big]_{x_1=0}, \label{var_action_0001}
\end{eqnarray}
which does not vanish with arbitrary choice of boundary condition.  However, it can be easily shown that the above vanishes for those boundary conditions which preserve $\mathcal{N}=1$ SUSY ( discussed in the last section).

When Dirichelt boundary condition is imposed on the scalar,
\begin{equation}
\Phi\Big|_{x^1=0}=0, \quad\quad \partial_0 \Phi\Big|_{x^1=0}=0, 
\end{equation}
it is easy to see that $\delta S$ vanishes.

When Neumann boundary  is imposed on the scalar and the theory is massless, the SUSY  conditions (\ref{relation_UF_1n}) and (\ref{nc1}) gives
\begin{eqnarray}
\bar{\epsilon} \partial_1 \Psi\Big|_{x^1=0}=0, \quad\quad \epsilon_-& =& \gamma^0 U_F \epsilon_+,  \label{massless_ncond001}
\end{eqnarray}
and the  boundary condition  (\ref{allowed_spinor_bc1}) yields
\begin{eqnarray}
\bar{\epsilon}\gamma^1\gamma^0 \Psi\Big|_{x^1=0} &=& -\Big[\epsilon^\dagger_+ \gamma^1 \Psi_- +\epsilon^\dagger_- \gamma^1\Psi_+\Big]_{x^1=0} \nonumber \\
&=& -\Big[\epsilon^\dagger_+ \gamma^1 \Psi_- +\epsilon^\dagger_+ U_F^\dagger \gamma^0 \gamma^1 U_F \gamma^0 \Psi_+\Big]_{x^1=0} \nonumber \\
&=& -\Big[\epsilon^\dagger_+ \gamma^1 \Psi_- +\epsilon^\dagger_+  \gamma^0 \gamma^1 U_F^\dagger U_F \gamma^0 \Psi_+\Big]_{x^1=0} \nonumber \\
&=& -\Big[\epsilon^\dagger_+ \gamma^1 \Psi_- +\epsilon^\dagger_+  \gamma^0 \gamma^1  \gamma^0 \Psi_+\Big]_{x^1=0} \nonumber  \\
&=& -\Big[\epsilon^\dagger_+ \gamma^1 \Psi_- -\epsilon^\dagger_+   \gamma^1  \Psi_+\Big]_{x^1=0} \nonumber\\
&=& 0.   \label{massless_ncond002}
\end{eqnarray}
Using (\ref{massless_ncond001}) and (\ref{massless_ncond002}) in (\ref{var_action_0001}) it is easy to check that in the massless case $\delta S$ vanishes. 

In the massive theory, when Neumann boundary condition is imposed on the scalar, along with (\ref{massless_ncond001}) and (\ref{massless_ncond002}), $U_F$ also satisfies $U_F^\dagger= U_F$. Owing to the last condition on $U_F$, it follows that
\begin{eqnarray}
\bar{\epsilon} \gamma^1 \Psi \Big|_{x^1=0} &=& \Big[\epsilon^\dagger_+ \gamma^0 \gamma^1 \Psi_+ +\epsilon^\dagger_-\gamma^0 \gamma^1\Psi_-\Big]_{x^1=0} \nonumber\\ 
 &=& \Big[\epsilon^\dagger_+ \gamma^0 \gamma^1 U_F \gamma^0 \Psi_- +\epsilon^\dagger_+ U_F^\dagger \gamma^0 \gamma^0 \gamma^1\Psi_-\Big]_{x^1=0} \nonumber\\ 
 &=& \Big[\epsilon^\dagger_+  U_F \gamma^0 \gamma^1\gamma^0 \Psi_- +\epsilon^\dagger_+ U_F^\dagger  \gamma^1\Psi_-\Big]_{x^1=0} \nonumber\\
&=& \Big[-\epsilon^\dagger_+  U_F \gamma^1 \Psi_- +\epsilon^\dagger_+ U_F^\dagger  \gamma^1\Psi_-\Big]_{x^1=0} \nonumber\\
&=& 0.   \label{massless_ncond003}
\end{eqnarray}
When (\ref{massless_ncond001}), (\ref{massless_ncond002}) and (\ref{massless_ncond003}) are substituted in (\ref{var_action_0001}), in the massive case also, we find, 
\begin{equation}
 \delta S = 0.
\end{equation}	
Therefore,these results are  consistent with the findings of the previous section.

\section{Edge states in (1+1)-dimension}

In these massive  $\mathcal{N}=1$ theories, for the choice of $\theta =(2n+1) \frac{\pi}{2}$ in  (\ref{U_F_gen_1d}), there are zero energy fermionic modes:
\begin{equation}
\Psi_e = G e^{b x_1} \left(\begin{array}{cc}
1\\
-(-1)^n\end{array}\right), \,\,\,\,\, b = (-1)^n m, \,\,\,\,\,\, n\in \mathbb{Z}.
\end{equation}  
 $G$ is the normalization constant. These modes are normalizable only for $m >0$ and $n=$even or $m<0$ and $n=$odd. If $|m|$ is sufficiently large, the zero modes are exponentially damped in the bulk $x^1 <0$ and are therefore localized near the boundary. For the scalar $\Phi$ however, there is no zero energy mode with Dirichlet  boundary condition. Thus the fermionic edge states, when present, are not paired with bosonic edge states. But such unpaired states do not break SUSY as they are zero energy modes and singlets under SUSY.  Consequently, when the boundary conditions are suitably chosen such that the edge states exist, the residual $\mathcal{N}=1$ supersymmetric theory has a fermionic ground state.

In the massless theory, such fermionic edge states do not exist because  there is no mass gap.

\section{(3+1)-dimensions}

In the full $(3+1)$-dimensional Minkowski spacetime, a theory with two complex scalars $\Phi_a$'s $(a=1,2)$ and a Dirac spinor $\Psi$ with the action
\begin{equation}
S=\int_{-\infty}^\infty d^4x \left(\mathcal{L}_{kin} + \mathcal{L}_{m} \right) 
\end{equation}
\begin{eqnarray}\left.
\begin{array}{lll}
 \mathcal{L}_{kin} &=& \frac{1}{2} \partial^\mu\Phi^\dagger_a \partial_\mu \Phi_a + i \bar{\Psi} \gamma^\mu \partial_\mu \Psi + \frac{1}{2}F_a^\dagger F^a , \\ \\
\mathcal{L}_{m} &=& m\left( \frac{i}{2} \Phi^\dagger _a F^a - \frac{i}{2} F_a^\dagger \Phi^a +\bar{\Psi}\Psi\right)
\end{array}\right.
\end{eqnarray}
is $\mathcal{N}=2$ supersymmetric  with a central charge $Z=P^\mu P_\mu$.
Here $F_a$'s are two complex scalar auxiliary fields which  are necessary to close SUSY off-shell. The non-zero central charge ensures that  particles with spin $> \frac{1}{2}$ are absent from the multiplet (for details see pages 150 -152 in \cite{sohnius}).
The supersymmetry transformations are
\begin{eqnarray}\label{supersymmetry_transfromations_1}
&&\delta \Phi_a = 2 \bar{\epsilon}^a \Psi, \\
\label{supersymmetry_transfromations_2}
&& \delta \Psi = -i \gamma^\mu (\partial_\mu \Phi_a) \epsilon^a -i F_a \epsilon^a,\\
\label{supersymmetry_transfromations_3}
&& \delta F_a =2 \bar{\epsilon}^a \gamma^\mu \partial_\mu \Psi,
\end{eqnarray}
where $ \epsilon^a$'s are a pair of constant 4-component spinors satisfying the reality condition:
\begin{equation}\label{symplectic_majorana}
\epsilon^1 = - C  \epsilon^{2\ast}, \quad \epsilon^2= C \epsilon^{1\ast}, \quad C= \gamma^1\gamma^3 \gamma^0.
\end{equation}

When a spatial  boundary $\partial M$ is inserted at $x^1=0$, in the resulting manifold $M$ the set of allowed uniform boundary conditions for the scalars $\Phi_a$'s is again given by (\ref{allowed_scalar_bc1}). But $U_B$ in this case is a $2\times 2$ matrix:
\begin{equation}
U_B=\left(\begin{array}{lll}
U_B^{11}& U_B^{12}\\
U_B^{21} & U_B^{22}
\end{array}\right), \,\,\,\,\,\,\,\,\,\, U_B^\dagger U_B=\mathbb{I}_{2\times 2}.
\end{equation}
Therefore, the local boundary conditions on the $\Phi_a$'s are 
\begin{equation}\label{sclar_bc_3d1}
 (\Phi_a+ i\partial_n \Phi_a)_{\partial M} = U_B^{ab} (\Phi_b-i \partial_n \Phi_b)_{\partial M}.
\end{equation}
For the choice $U_B^{ab}=-\delta^{ab}$ and $U_B^{ab}=\delta^{ab}$, we get the Diraichlet and Neumann boundary conditions respectively. For $U^{ab}_B \neq \pm \delta^{ab}$, we get the another type of boundary conditions: $\partial_n \Phi_a\Big|_{\partial M} = \kappa^{ab} \Phi_b\Big|_{\partial M}$ where $\kappa^{ab}=i  \left((\mathbb{I}+U_B)^{-1}(\mathbb{I}-U_B)\right)^{ab}$.

On the spinor $\Psi$, again the boundary conditions (\ref{allowed_spinor_bc1}) and (\ref{cond_953}) are imposed. But unlike (\ref{cond_952}) in the (1+1)-dimensional massless case, (\ref{cond_953})  involves the tangential derivates of $\Psi$ at the boundary.

The supersymmetry transformation at the boundary  must obey
\begin{equation}
\delta (\Phi_a+i \partial_n \Phi_a)_{\partial M} = U_B^{ab} \delta (\Phi_b-i \partial_n \Phi_b)_{\partial M}.
\end{equation}
Using (\ref{supersymmetry_transfromations_1}), the above yields
\begin{equation}\label{susy_3d_bc}
\left[(\delta^{ab}-U_B^{ab}) \bar{\epsilon} ^b \Psi + i(\delta^{ab} +U_B^{ab}) \bar{\epsilon} ^b \partial_n \Psi\right]_{\partial M} =0.
\end{equation}

\textbf{Dirichlet boundary condition:}
If we impose Dririchlet boundary conditions on both the scalars: $U_B=-\mathbb{I}_{2\times 2}$, then (\ref{susy_3d_bc}) and (\ref{allowed_spinor_bc1}) give (similar to (1+1)-dimensional case)
\begin{equation}\label{relation_UF_3d}
\epsilon^{a}_+ =- U_F^\dagger \gamma^0 \epsilon^a_-.
\end{equation}
Because $\epsilon_a$'s are  constant spinors, the above is true not only on the boundary but also in the bulk.
Further, using Dirichlet boundary conditions on $\Phi_a$'s and (\ref{relation_UF_3d}), it is easy to check that on the  boundary $\partial M$ (similar to (1+1)-dimensional case)
\begin{equation}\label{proof_234}
\delta \Psi _+ = U_F \gamma^0 \delta \Psi _+, \quad \delta (H_D \Psi)_+ = U_F \gamma^0 \delta(H_D \Psi) _+.
\end{equation}
Thus Dirichlet boundary conditions on both scalars are compatible with supersymmetry transformations. But as the $\epsilon^a$'s are related by (\ref{relation_UF_3d}), the theory is only $\mathcal{N}=1$ supersymmetric.

The closure of the SUSY algebra is governed by 
\begin{equation}
[\delta_\epsilon ,  \delta_\eta] = 2i \bar{\epsilon}_a \gamma^\mu \eta^ a \partial_\mu + 2i\bar{\epsilon}_a \eta^a \delta_Z
\end{equation}
where $\delta _Z$ gives the action of the central charge on the fields. 
The seond term in the above vanishes in the massless case and in the massive case it can be absorbed by rescaling the momenta, in a similar fashion as in the (1+1)-dimensional case (see (\ref{modified_algebra})). 

\textbf{Neumann and Robin boundary conditions:}
It is easy to check that if we impose Neumann or Robin-type boundary condition on either or both of the scalar fields $\Phi_a$'s , then (\ref{allowed_spinor_bc1}), (\ref{cond_953}) and (\ref{susy_3d_bc}) cannot be satisfied. \textit{So such boundary conditions on scalars are not consistent with supersymmetry and  SUSY is completely broken.}

\textit{However, if we insert a  boundary $\partial M$ with Dirichlet boundary conditions on scalars, the theory can still be supersymmetric. The theory with $\mathcal{N}=2$ SUSY  breaks to an  $\mathcal{N}=1$ supersymmetric theory for every allowed $U_F$. 
For any other choice of boundary conditions, SUSY is completely broken.}

\section{Edge states in (3+1)-dimensions}

In the following we investigate the possibility of existence of edge states in  theories which have residual $\mathcal{N}=1$ SUSY.
For simplicity, let us consider the region $x_3 \leq 0$ as the $(3+1)$-dimensional flat manifold $M$. On the boundary  plane $x_3=0$, the direction of the outward normal is $\hat{n} =(0,0,1)$.
We choose the $\gamma$-matrices in the representation
\begin{equation}\label{gamma_3d_rep}
\gamma^\mu =\left(\begin{array}{llll}
0 & \sigma^\mu \\
\bar{\sigma}^\mu & 0
\end{array}\right), \quad \sigma^\mu=(1, \sigma^i),\quad \bar{\sigma}^\mu=(1, -\sigma^i).
\end{equation}
In this case $U_F$ satisfies
\begin{equation}
[U_F, \gamma^0 \gamma^3]=0, \,\,\,\,\,\, U_F^\dagger U_F= \mathbb{I},\,\,\,\,\, CU_F^T-U_F^\dagger C=0.
\end{equation}
The last condition in the above is imposed by (\ref{symplectic_majorana}) and (\ref{relation_UF_3d}). Therefore, the most general $U_F$ in this case is given by (detailed derivation is given in appendix \ref{appendix_gen_UF})
\begin{equation}\label{3d_UF}
U_F=\left(
\begin{array}{cccc}
v_{1} &0&0&v_{2}\\
0&u_{1}^{\ast} &-u_{2}^{\ast}&0\\
0&u_{2}&u_{1}&0\\
-v_{2}^{\ast}&0&0& v_{1}^{\ast}
\end{array}\right), \,\, u_1, u_2,v_1, v_2 \in \mathbb{C},
\end{equation}
with $|u_1|^2+ |u_2|^2=1$ and $|v_1|^2+ |v_2|^2=1$.
\newline 
i) {\textit{Massless case:}}
In the massless case, 
if
$
u_{1}=0$, $Re (u_{2})\neq 0$ and $ Im (u_{2})\neq 0$ are chosen
in   (\ref{3d_UF}),  there exist two zero energy edge localized states for arbitrary $b>0$:
\begin{eqnarray}
&&\hspace{-1.1cm}\Psi^0_{e_1}=A_k \left(
\begin{array}{cccc}
1\\
 -u_{2\ast}\\
0 \\
0
\end{array}\right)e^{b x^3+i k_1 x^1+ik_2 x^2}, \\
\hspace{-2.2cm}\textrm{with } \,\,\, k_1= b\, Im (u_{2}), \quad  k_2= b\, Re (u_{2}),  \textrm{ and}\nonumber
\end{eqnarray}
\begin{eqnarray}
&&\hspace{-1.1cm}\Psi^0_{e_2}=D_k \left(
\begin{array}{cccc}
0\\
0\\
 u_{2}\\
1
\end{array}\right)e^{b x^3+i k_1 x^1+ik_2 x^2}, \\
\hspace{-2.4cm}\textrm{with }\,\,\,  k_1=- b\, Im (u_{2}), \quad k_2= -b\, Re (u_{2}).\nonumber 
\end{eqnarray}
 $A_k$ and $D_k$ are normalization constants. As $\Psi_{e_1}^{0\dagger} \Psi_{e_2}^0=0$, these two modes are linearly independent.
For sufficiently large $b$, these modes are localized near the edge and are exponentially damped in the bulk.

For this choice of $u_1$ and $u_2$, there does not exist any other normalizable  zero energy edge state. 

ii) {\textit{Massive case:}}
If we choose $Re( u_1)=0$, $Im(u_1) \neq 0$, $Re( u_{2}) \neq 0$ and  $Im( u_{2})\neq0$
in  (\ref{3d_UF}), there exist either of the following two zero-energy states:
\newline 
a) For $Im(u_1)<0$,
\begin{equation}
\Psi_{e_1}^m= A_k\left(
\begin{array}{cccc}
1\\ 
-u_2^\ast\\ 
u_1 \\ 
0
\end{array}\right)e^{b x^3+i k_1 x^1+ik_2 x^2},
\end{equation}
with $ 
b= -\frac{m}{Im(u_1)}$, $ k_1 = b\, Im (u_{2})$ and $ k_2 = b\, Re( u_{2}). $
($A_k$ is the normalization constant.) \\
b) For $Im(u_1)>0$,
\begin{equation}
\Psi_{e_2}^m= D_k\left(
\begin{array}{cccc}
0\\ 
u_1^\ast
\\ 
u_2
\\ 
1
\end{array}\right)e^{b x^3+i k_1 x^1+ik_2 x^2},
\end{equation}
with $b= \frac{m}{Im(u_1)}$, $  k_1 = -b\,Im (u_{2})$ and $ k_2 = -b\,Re( u_{2})$.
($D_k$ is the normalization constant.)
If $m$ is very large and/or $|Im(u_1)|$ is very small, these states are exponentially damped in the bulk and are localized near the edge. 

 For these choices of $u_{1}$ and $u_2$, there does not exist any other normalizable edge state. 

For a scalar field obeying Dirichlet boundary conditions, there are no zero-energy modes of the Laplacian (for details see appendix \ref{appendix_scalar mode}). On the other hand, it is possible to 
choose boundary conditions for the fermion such that there exist fermionic zero modes. In such a situation, the ground state is made up of a 
fermion but no boson. This however does not break supersymmetry, precisely because it is a zero-energy state.

If such fermionic  edge states exist, it should be possible to experimentally detect them in condensed matter systems, especially in the supersymmetric phase of superconductors.

\section{Conclusions and Discussions}
We have shown that when spatial boundaries are introduced in an $\mathcal{N}=2$ supersymmetric theory, SUSY is broken. For only a few boundary conditions  can SUSY be partially preserved.  For other boundary conditions, SUSY is completely broken. 
As we have shown, it is possible to extend our analysis to any spacetime dimension.  
Though we have considered only flat boundaries  for the simplicty of our analysis, it is not difficult to see that the results will be true in general, for any curved boundary.  Also, the above analysis is valid not only for $\mathcal{N}=2$, but also for any $\mathcal{N}=$ even supersymmetric theory. In our analysis, we considered free theories. However, one might consider interactions as well and in that case, it is not difficult to convince onefself that the results should be in consistency with \cite{Gaiotto:2008sa}.
We expect that edge states in these interacting theories (more realistic  in the context of say, a superconductor) exist in a similar fashion. Nonetheless, 
the details of the properties of these states needs to be studied.

The presence of the edge localized fermions as ground states of these supersymmetric theories is important in the context of  systems like topological superconductors. For example, these fermions localized on the boundary will contribute to the Meissner effect of the  superconductor and thus experimental verification of these fermions localized in the boundary is possible.

\section*{Acknowledgements}
We would like to thank Romesh Kaul and Subroto Mukerjee for illuminating discussions and suggestions. APB thanks Centre for High Energy Physics, Indian Institute Science, Bangalore and especially Sachin Vaidya for hospitality during the course of this work. He also thanks Andres Reyes at the Universidad de los Andes for hosting him as the Sanford Professor and for warm hospitality when this work was being completed. The work of MA  has been partially supported by  Spanish DGIID-DGA grant 2014-E24/2 and  Spanish
MICINN grants FPA2012-35453 and CPAN-CSD2007-00042.

\begin{appendices}

\section{The Most General $U_F$}\label{appendix_gen_UF}

In the (3+1)-dimensionsal manifold $M = \{x_3\leq 0\}$, $U_F$ is of the form
\begin{eqnarray}
U_F=\left(
\begin{array}{lll}
U_1 & U_2 \\
U_3 & U_4
\end{array}
\right),
\end{eqnarray}
where $U_1,\,U_2,\,U_3$ and $U_4$ are $2\times 2$ matrices.  
Also $U_F$ satisfies 
\begin{equation}
 [U_F,\gamma^0\gamma^3]=0, \quad CU_F^T-U_F^\dagger C=0, \quad  U_F^\dagger U_F=\mathbb{I}.
\end{equation}

With the chioice of $\gamma$-matrices as in (\ref{gamma_3d_rep}), 
the condition $ [U_F,\gamma^0\gamma^3]=0$ yields
\begin{eqnarray}
\begin{array}{lll}
&& \left(
\begin{array}{cc}
-[U_1,\sigma^3] & \{U_2,\sigma^3\} \\
-\{U_3,\sigma^3\} & [U_4,\sigma^3]
\end{array}
\right)=0.
\end{array}
\end{eqnarray}
Hence,
\begin{eqnarray}
\begin{array}{lllll}
[U_1,\sigma^3]=0, &\quad\quad \{U_2,\sigma^3\}=0, &\quad\quad
\{U_3,\sigma^3\}=0, &\quad\quad [U_4,\sigma^3]=0.
\end{array}
\end{eqnarray}
Therefore, the most general  $U_1$ and $U_4$ are
\begin{eqnarray}\label{diag_U1_U4}
\begin{array}{lll}
U_1=\left(
\begin{array}{cc}
v_1 & 0 \\
0 & v_4
\end{array}
\right), \quad U_4=\left(
\begin{array}{cc}
u_1 & 0 \\
0 & u_4
\end{array}
\right),\quad\quad u_1,v_1,u_4,v_4\in\mathbb{C},
\end{array}
\end{eqnarray}
and the most general $U_2$ and $U_3$ are
\begin{eqnarray}
\begin{array}{lll}
U_2=\left(
\begin{array}{cc}
0 & v_2 \\
v_3 & 0
\end{array}
\right),
\end{array}
\quad
\begin{array}{lll}
U_3=\left(
\begin{array}{cc}
0 & u_2 \\
u_3 & 0
\end{array}
\right),\quad\quad u_2,v_2,u_3,v_3\in\mathbb{C}.
\end{array}
\end{eqnarray}

From (\ref{diag_U1_U4}), it is easy to see that  
\begin{eqnarray}
U_1^\dagger = U_1^\ast, \quad U_1^T = U_1, \quad U_4^\dagger = U_4^\ast, \quad U_4^T = U_4.
\end{eqnarray}
Hence, the condition $CU_F^T-U_F^\dagger C=0 $ leads to 
\begin{eqnarray}
\begin{array}{lll}
&& \sigma^2U_2^T-U_3^\dagger\sigma^2=0,\quad  \sigma^2U_4-U_1^*\sigma^2=0.
\end{array}
\end{eqnarray}
These yield
\begin{eqnarray}
v_4 = u_1^\ast, \quad u_4= v_1^\ast, \quad  v_3= - u_2^\ast, \quad u_3= - v_2^\ast.
\end{eqnarray}
Therefore, 
\begin{equation}
U_F=\left(
\begin{array}{cccc}
v_{1} &0&0&v_{2}\\
0&u_{1}^{\ast} &-u_{2}^{\ast}&0\\
0&u_{2}&u_{1}&0\\
-v_{2}^{\ast}&0&0& v_{1}^{\ast}
\end{array}\right), \,\, u_1, u_2,v_1, v_2 \in \mathbb{C}.
\end{equation}

The unitarity condition $U_F^\dagger U_F=\mathbb{I}$ gives
$$|v_1|^2 +|v_2|^2=1, \quad |u_1|^2 +|u_2|^2=1.$$

\section{Zero Modes of the Scalar Field}\label{appendix_scalar mode}
In the (3+1)-dimensional manifold $M=\{x_3 \leq 0\}$, the zero modes of the scalar field are
\begin{eqnarray}
&&(- \nabla^2 +m^2) \Phi(x) =0, \\
&&\Phi(x) = (A_k e^{bx_3} + B_k e^{-bx_3}) e^{i(k_1 x_1 + k_2 x_2)},
\end{eqnarray}
with
\begin{equation}
b^2 = k_1^2 + k_2^2.
\end{equation}
As $x_3 \rightarrow -\infty$,  $\Phi $ must go to zero. Hence, $B_k=0$ and 
\begin{eqnarray}
&&\Phi (x)= A_k e^{bx_3+i(k_1 x_1 + k_2 x_2)}. 
\end{eqnarray}
Imposing the Dirichlet boundary condition at $x_3=0$: 
\begin{equation}
\Phi(x) \Big|_{\partial M} =0,
\end{equation} 
 yields
\begin{equation}
A_k =0. 
\end{equation}
Hence, there are no zero energy scalar modes. 

\end{appendices}

\end{document}